\documentclass[twocolumn]{aastex631}

\newcommand{\Msun}{\rm M_\odot}
\newcommand{\kpc}{\,\rm kpc}
\newcommand{\re}{r_{\rm e}}
\newcommand{\fdm}{f_{\rm DM}}
\newcommand{\fbar}{f_{\rm baryon}}
\newcommand{\Mgas}{M_{\rm gas}}
\newcommand{\Mdm}{M_{\rm DM}}
\newcommand{\Mbar}{M_{\rm baryon}}
\newcommand{\Mdyn}{M_{\rm dyn}}

\usepackage{float,subfloat}

\begin{document}

\title{An early dark matter-dominated phase in the assembly history of Milky Way-mass galaxies suggested by the TNG50 simulation and JWST observations}

\author[0000-0002-2380-9801]{Anna de Graaff}\affiliation{Max-Planck-Institut f\"ur Astronomie, K\"onigstuhl 17, D-69117, Heidelberg, Germany}\email{degraaff@mpia.de}
\author[0000-0003-1065-9274]{Annalisa Pillepich}\affiliation{Max-Planck-Institut f\"ur Astronomie, K\"onigstuhl 17, D-69117, Heidelberg, Germany}%\email{pillepich@mpia.de}
\author[0000-0003-4996-9069]{Hans-Walter Rix}\affiliation{Max-Planck-Institut f\"ur Astronomie, K\"onigstuhl 17, D-69117, Heidelberg, Germany}%\email{rix@mpia.de}

\begin{abstract}
\noindent Whereas well-studied galaxies at cosmic noon are found to be baryon-dominated within the effective radius, recent JWST observations of $z\sim6-7$ galaxies with stellar masses of only $M_*\sim10^{8-9}\,\Msun$ surprisingly indicate that they are dark matter-dominated within $\re\approx 1\kpc$. Here, we place these high-redshift measurements in the context of the TNG50 galaxy formation simulation, by measuring the central  (within $1\kpc$) stellar, gas, and dark matter masses of galaxies in the simulation. The central baryon fraction varies strongly with galaxy stellar mass in TNG50, and this $M_*$-dependence is remarkably constant across $0<z<6$: galaxies of low stellar mass ($M_*\sim10^{8-9}\,\Msun$) are dark matter-dominated, as $\fbar(<1\kpc)\sim0.25$\,.  At $z=6$, the baryonic mass in the centers of low-mass galaxies is largely comprised of gas, exceeding the stellar mass component by a factor $\sim4$. We use the simulation to track the typical evolution of such low-mass dark matter-dominated galaxies at $z=6$, and show that these systems become baryon-dominated in their centers at cosmic noon, with high stellar-to-gas mass ratios, and grow to galaxies of $M_*\sim10^{10.5}\,\Msun$ at $z=0$. Comparing to the dynamical and stellar mass measurements from observations at
high redshifts, these findings suggest that the inferred star formation efficiency in the early Universe is broadly in line with the established assumptions for the cosmological simulations. Moreover, our results imply that the JWST observations may indeed have reached the early, low-mass regime where the central parts of galaxies transition from being dark matter-dominated to being baryon-dominated. 
\end{abstract}

\section{Introduction} \label{sec:intro}

In the $\Lambda$CDM framework, galaxy formation proceeds through the collapse of dark matter haloes, after which baryons can cool and form stars \citep[e.g.][]{White1978}. Galaxies grow through the accretion of (cool) gas and subsequent star formation, as well as by merging with nearby systems \citep[e.g.,][]{Dekel2009a,Oser2010}. However, the details of these processes, their relative importance as a function of cosmic time, and the interplay between the dark and baryonic matter are still unclear.

Constraints on the mass budget, i.e. the relative contribution of dark matter, gas and stellar mass, can offer insight into the mass assembly histories of galaxies. In the local Universe, moderately massive galaxies ($M_*\sim10^{10-11}\,\Msun$) are baryon-dominated near their centers ($\sim1\kpc$), but contain substantial dark matter mass within the effective radius ($\fdm(<\re)\sim 40-70\%$ for $\re\sim5-10\kpc$; \citealt{Martinsson2013}). Studies of ionized gas kinematics at $z\sim1-3$ paint a very different picture, as these find an increase in the baryonic mass fraction within $\re$ toward higher redshift for galaxies of $M_*\gtrsim 10^{9.5}\,\Msun$, with galaxies appearing almost entirely baryon-dominated at $z\sim2$ \citep{Wuyts2016,Price2016,Price2020,Genzel2017,Genzel2020,Nestor2023}{, although recent work suggests this may not be the case for all galaxies at this epoch \citep{Sharma2023}}. For very massive galaxies at $z\sim2$ \citet{Genzel2017,Genzel2020} propose {that such high baryon fractions} may imply that the dark matter distributions are cored instead of cuspy, which may be caused by a very rapid formation process coupled with stellar and black hole feedback.

The James Webb Space Telescope (JWST) enables the measurement of galaxy kinematics at even higher redshifts and also lower masses, thereby probing the possible progenitors of more massive galaxies at cosmic noon. In \citet{deGraaff2023} we used JWST/NIRSpec observations to measure the ionized gas kinematics for six small ($\re\sim1\kpc$) galaxies at $z\sim6-7$ of low stellar mass, $M_*\sim10^{7-9}\,\Msun$. Interestingly, the dynamical masses inferred from the ionized gas kinematics exceed the stellar masses by a factor $\sim 30$, and the baryonic masses by a factor $\sim 3$, which may imply surprisingly high dark matter fractions of $\fdm(<\re)\approx0.7$. However, these JWST results are still subject to large observational and systematic uncertainties, raising the question whether these implied dark matter fractions are realistic, and whether these are compatible with observations at intermediate redshifts.

In this Letter, we use the cosmological magneto-hydrodynamical galaxy simulation known as TNG50 \citep{Nelson2019,Pillepich2019}, and part of the IllustrisTNG project\footnote{\url{www.tng-project.org}}, to extract theoretical expectations for the baryonic mass fraction of galaxies on small spatial scales as a function of redshift and stellar mass. For relatively massive galaxies in the lower-redshift Universe ($z\lesssim2-3$), previous work within IllustrisTNG has provided an overall quantification of the dark matter fraction \citep{Lovell2018} and a direct comparison to observations \citep{Uebler2021}.  Here, we use TNG50 to focus on low stellar mass galaxies at $z=6$, in order to compare the simulation expectations to the observational results from JWST at $z\sim6-7$. By tracing the descendant population in the simulation, we check whether the JWST results can be linked to, and are consistent with, ground-based measurements at cosmic noon. We use physical lengths when specifying distances and sizes throughout, unless otherwise stated.

\section{TNG simulations}\label{sec:data}

We base our main analysis on the TNG50 simulation \citep{Nelson2019,Pillepich2019}, the highest resolution simulation in the IllustrisTNG project \citep[TNG hereafter;][]{Marinacci2018,Naiman2018,Nelson2018a,Pillepich2018b,Springel2017}. TNG50 assumes a $\Lambda$CDM cosmology with cosmological parameters from the \citet{Planck2016}, and uses the moving mesh code AREPO \citep{Springel2010} and the IllustrisTNG galaxy formation model \citep{Weinberger2017,Pillepich2018a} to simulate several thousand galaxies simultaneously within a cosmological volume of $\sim50^3$ comoving Mpc$^3$. 

With a baryonic mass resolution of $m_{\rm baryon}=8.5\times10^4\,\Msun$, a dark matter particle resolution of $m_{\rm DM}=4.5\times10^5\,\Msun$ and a Plummer-equivalent gravitational softening length of $\epsilon^{z=0}_{\rm DM, *}=0.29\kpc$ at $z=0$, TNG50 has both the large volume and the high resolution needed to robustly evaluate the average mass profiles of galaxies on scales of $\sim1 \kpc$ across a wide range in redshift and stellar mass. 
To test the convergence of our results, we also use the TNG100 simulation \citep{Nelson2019a}, which employs the same galaxy formation model, but for a larger volume of $\sim100^3$ comoving Mpc$^3$ and lower resolution ($m_{\rm baryon}=1.4\times10^6\,\Msun$; $m_{\rm DM}=7.5\times10^6\,\Msun$; $\epsilon^{z=0}_{\rm DM, *}=0.74\kpc$).

Haloes and subhaloes are identified using the Friends-of-Friends (FoF) \citep{Davis1985} and \textsc{subfind} \citep{Springel2001} algorithms, respectively. We define the effective radius as the 3D stellar half-mass radius of the subhalo: $\re \equiv r_{1/2,*}$. Because we trace a wide range in mass and redshift (and hence galaxy size), we define the total stellar mass as the stellar mass enclosed within a spherical aperture of twice the half-mass radius, $M_*\equiv M_*(<2\re)$, instead of using a fixed aperture to compute stellar masses.

We define our primary sample by initially selecting central galaxies identified by \textsc{subfind} at $z=6$ (snapshot 13) with a stellar mass $M_*>10^7\,\Msun$, which are resolved by $>100$ stellar particles. From this sample of 3926 systems we further select the subset of 381 galaxies with $10^8\,\Msun < M_*<10^9\,\Msun$ (i.e. resolved by $>1000$ stellar particles), which we use to trace the descendant population and to compare with observations across different redshifts.

\section{Mass budget at $z\sim6$}\label{sec:mass_budget_z6}

We start assessing the central mass budget of high-redshift galaxies by looking at our sample of TNG50 central galaxies at $z=6$, which is the median redshift of the JWST measurements by \citealt{deGraaff2023}, and measuring the baryonic mass fraction in a (3D) spherical aperture of the stellar half-mass radius:
\begin{equation}
    \fbar(<\re) = \frac{M_*(<\re)+\Mgas(<\re)}{M_*(<\re)+\Mgas(<\re)+\Mdm(<\re)}\,,
\label{eq:fbar}
\end{equation}
where $\Mgas$ ($\Mdm$) is the mass of all gas cells (dark matter particles) within the aperture.

\begin{figure}
    \centering
    \includegraphics[width=\linewidth]{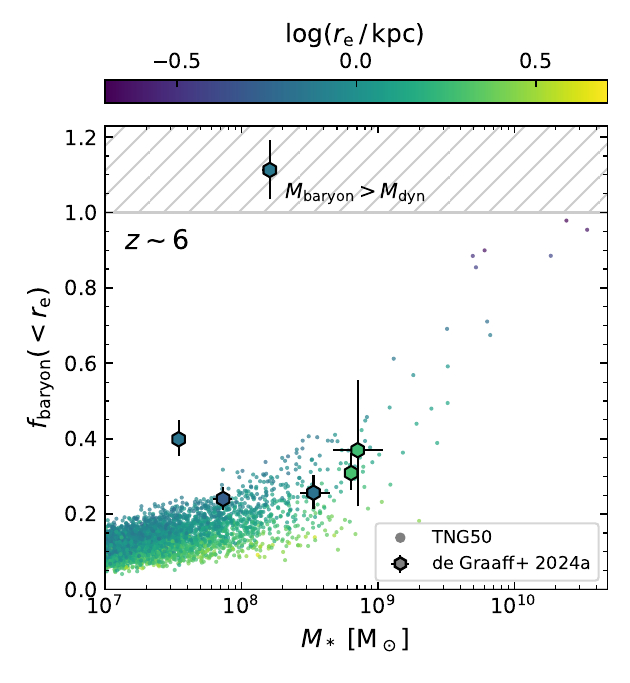}
    \caption{Baryonic mass fraction within the stellar half-mass radius (TNG50) and the emission line half-light radius (JWST; \citealt{deGraaff2023}) of galaxies vs. total stellar mass. Throughout, by baryonic fraction we mean the ratio between the mass in gas and stars and the total mass of the galaxy (see Eq.~\ref{eq:fbar}). The TNG50 simulation predicts low baryon fractions for $M_*<10^9\,\Msun$ and a sharp increase in $\fbar$ toward higher masses. The half-mass radii are small ($\sim 1\kpc$ in physical units) and, at fixed stellar mass, are anti-correlated with the baryon fraction. With the exception of one source, which may have high $\fbar$ due to an underestimated dynamical mass, the baryon fractions measured at $z\sim6$ with JWST are in good agreement with the simulated galaxies. }
    \label{fig:z6_comp}
\end{figure}

Figure~\ref{fig:z6_comp} shows $\fbar(<\re)$ versus the total stellar mass in TNG50, color-coded by $\re$: there is a strong dependence of $\fbar$ on stellar mass. Galaxies of low stellar mass ($M_*\lesssim10^{8.5}\,\Msun$) have baryon fractions in the range $\fbar(<\re)\sim0.1-0.3$ and thus are dark matter-dominated. Above $M_*\sim10^{8.5}\,\Msun$, $\fbar(<\re)$ increases rapidly in TNG50, and the most massive systems ($M_*\sim10^{10}\,\Msun$) are baryon-dominated. {At fixed stellar mass, the scatter in $\fbar(<\re)$ among individual galaxies is modest ($\sigma(\fbar|M_*=10^8\,\Msun)\approx0.05$), and covariant with $\re$, so that larger effective radii imply lower values of $\fbar$.}
%Moreover, $\fbar(<\re)$ is covariant with $\re$ at fixed stellar mass, so that larger effective radii imply lower values of $\fbar$. 

These simulation results are in remarkably good agreement with the measurements from \citet{deGraaff2023}, inferred from JWST observations of low-mass galaxies ($M_*\sim10^{7-9}\,\Msun$) at $z\sim6-7$. The observational estimates of the baryonic mass were obtained from spectral energy distribution (SED) modeling to low-resolution spectroscopy, while the total mass estimates were inferred dynamically from measurements of ionized gas kinematics. We emphasize that these observed estimates differ conceptually in important ways from those for the simulated baryon fractions. For example, the observed effective radius represents the \emph{projected half-light radius of the emission line} instead of a half-mass radius, although we find that these are $\sim 1\kpc$ in both simulated and observed galaxies. The stellar masses were inferred from SED modeling and the cold-gas mass is empirically inferred from the star formation rate (SFR); both observational estimates carry significant systematic uncertainties {that are not reflected in the error bars shown.

For a complete discussion of the systematic uncertainties in the different mass estimates we refer the reader to \citealt{deGraaff2023}, but we provide a brief summary here. The stellar mass in particular can be underestimated by as much as a factor 10 in extreme scenarios where only the integrated light of the galaxy is considered in the SED fitting and a luminous young stellar population outshines the underlying population of older stars in the galaxy \citep[e.g.][]{Gimenez2023}. However, the spatially-resolved analysis of one of the galaxies shown in Figure~\ref{fig:z6_comp}, with an SED that is representative of the full $z\sim6$ sample, yields an integrated stellar mass that is consistent with the mass inferred from SED fitting to the integrated light \citep{Baker2023}. The inferred gas masses are also uncertain, as these were obtained from the conversion of a star formation surface density under the assumption of a constant star formation efficiency (\citealt{Kennicutt1998}; for further discussion see also Section~\ref{sec:discussion}), whereas we may expect this efficiency to increase toward higher redshift \citep{Tacconi2020}. The star formation efficiency in the low-mass, high-redshift regime is still poorly constrained observationally, but we note that \citet{Price2020} found that for galaxies at cosmic noon the gas masses estimated using the empirical relations by \citet{Kennicutt1998} and \citet{Tacconi2020} differ by only 0.13\,dex. Such a systematic difference (i.e a factor 1.3 in the inferred $\fbar$) is still consistent within the scatter seen among the individual galaxies in TNG50.} 

Finally, we note that observations are subject to projection effects that can lead to an underestimated circular velocity. This may for example explain the data point for which the baryonic mass exceeds the total dynamical mass ($\Mbar>\Mdyn$). In light of these different issues, the broad agreement found between the observations and the TNG50 data shown in Figure~\ref{fig:z6_comp} is striking.

\begin{figure*}
    \centering
    \includegraphics[width=0.88\linewidth]{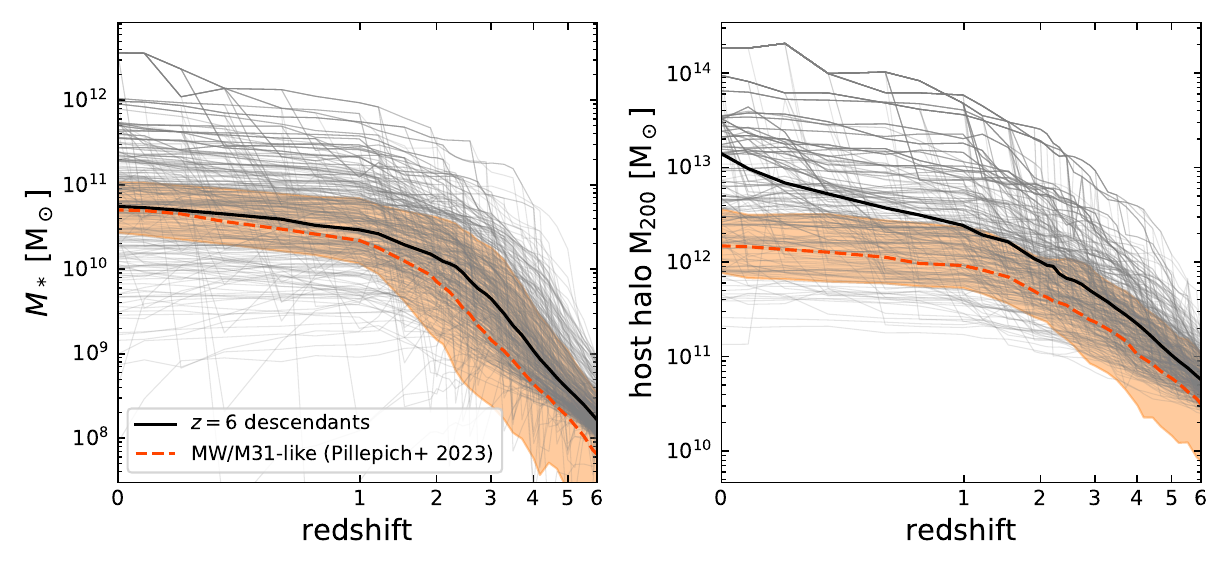}
    \caption{Mass assembly histories of low-mass TNG50 galaxies, selected to have $10^8\,\Msun < M_*<10^9\,\Msun$ at $z=6$. Although all %(381) 
    $z=6$ galaxies are selected to be centrals in their dark matter haloes, many systems merge to form massive centrals or massive satellites in groups and clusters at lower redshifts. To construct the median mass history (solid black lines), we consider only unique subhaloes in each snapshot. The median stellar mass of the $z=6$ descendant population increases 100-fold by $z=2\,$; at $z=0$ the median stellar mass of the population equals that of the Milky Way, but with large scatter (0.6\,dex). In orange we show the mass assembly histories (median and 5-95 percentiles) for the population of TNG50 MW and M31 analogs from \citet{Pillepich2023}, which were selected as analogs not only by stellar mass, but also by environment and stellar morphology. We find that 50 out of 213 galaxies of the $z=6$ descendant population are indeed MW or M31 analogs at $z=0$.}
    \label{fig:trees}
\end{figure*}

\section{Redshift evolution of the baryon and dark matter fraction in TNG50}\label{sec:redshift_evo}

Next, we use these same simulations to trace the mass assembly and mass budgets of $z=6$ galaxies to lower redshifts. To do so, we select 381 galaxies within the stellar mass range $10^8\,\Msun < M_*<10^9\,\Msun$ (at $z=6$) and use the merger trees identified with \textsc{sublink} \citep[i.e., the DM-only merger trees from][]{Rodriguez2015} to link progenitors and descendants. 

\subsection{Mass assembly histories}\label{sec:trees}

Although all galaxies were selected to be the central galaxy in their FoF halo at $z=6$ , at lower redshift these systems may fall into larger haloes and hence become satellite galaxies. Moreover, the $z=6$ galaxies are not necessarily the main progenitor of their $z=0$ descendants. Therefore, we forward trace the merger history for a subhalo selected at $z=6$ by iteratively searching for its next descendant; this is straightforward if the subhalo at a given snapshot is the main progenitor of the descendant in the following snapshot. If this is not the case (usually due to a merger with a more massive subhalo, either from the initial sample or a massive system not in the initial selection), we trace the subsequent evolution of the product of the merger. This implies that from the 381 galaxies at $z=6$, only 213 descendants remain by $z=0$. 
When calculating population statistics, we only consider unique descendant galaxies, i.e. we do not double count objects that have merged to form a single system at $z<6$.

In Figure~\ref{fig:trees} we illustrate the mass assembly histories of the $z=6$ descendant population through the redshift evolution of the stellar and the host halo mass ($M_{200}$). Thin gray lines show individual mass histories, and the black line shows the evolution of the ensemble median. In TNG50, the low-mass galaxies at $z=6$ typically grow to $10^{10}\,\Msun$ by $z=2$, and are therefore comparable in stellar mass to the objects that are typically studied kinematically at cosmic noon \citep[e.g.][]{Wuyts2016,Price2016,Price2020,Wisnioski2015,Wisnioski2019}. By $z=0$ the median stellar mass is $10^{10.5}\,\Msun$ -- comparable  to the stellar mass of the Milky Way \citep[e.g. see][]{Licquia2015,BlandHawthorn2016} --  although with an ensemble scatter of 0.6\,dex. In contrast, the typical descendant halo mass ($\sim10^{13}\,\Msun$) far exceeds the virial halo mass of the Milky Way ($\sim10^{12}\,\Msun$).

For comparison, we show in Figure~\ref{fig:trees} the sample of Milky Way and Andromeda (hereafter MW and M31) analogs from \citet{Pillepich2023}, who selected 198 TNG50 galaxies at $z=0$ with stellar masses, morphologies and 1\,Mpc scale environment similar to the MW and M31. At fixed redshift, these analogs typically have slightly lower stellar masses and significantly lower host halo masses than the sample of $z=6$ descendants selected in this work. However, we note that there is overlap between the two populations, as 50 galaxies of the 213 descendants in our selection are also in the MW/M31 analog sample of \citet{Pillepich2023}.

\begin{figure*}
    \centering
    \includegraphics[width=0.95\linewidth]{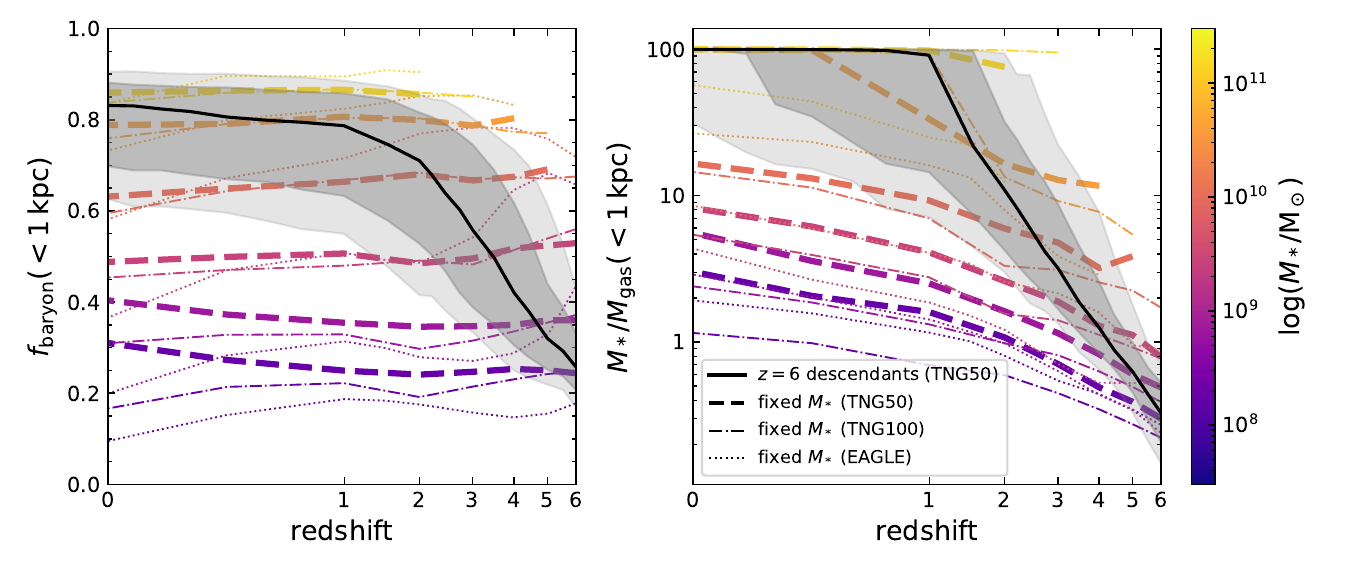}
    \caption{Evolution in the baryonic mass fraction (left) and stellar-to-gas mass ratio (right) within a fixed spherical aperture of radius 1\,kpc according to the IllustrisTNG and EAGLE cosmological hydrodynamical galaxy simulations. Colored lines show the median $\fbar$ and $M_*/\Mgas$ ratio within narrow stellar mass ranges for central galaxies in the TNG50 (dashed) and TNG100 (dash-dotted) simulations and, for comparison, also the EAGLE Ref-L100N1504 simulation (dotted). Solid black lines show the median time evolution of the descendant population of low-mass ($M_*\sim10^{8-9}\,\Msun$) galaxies selected from TNG50 at $z=6$, with shaded regions indicating the 16-84 and 5-95 percentiles. As the $z=6$ galaxies grow in stellar mass, $\fbar$ and $M_*/\Mgas$ increase rapidly, resulting in stellar mass-dominated centers by $z\sim2$. }
    \label{fig:fdm_aper}
\end{figure*}

\subsection{The central mass component budget: \\stars, gas and dark matter}\label{sec:apertures}

As mentioned in Section~\ref{sec:mass_budget_z6}, the observations and simulations differ significantly in their measurement methodology for all components of the overall mass budget: stars, gas and dark matter. Notably, the effective radii represent different physical quantities, one being based on stellar mass and the other on emission line extent. Moreover, although the mass-size relation in TNG is broadly in agreement with observations at $z\leq4$ and $M_*>10^9\,\Msun$ \citep{Pillepich2019}, it is still unclear whether this holds at higher redshifts and lower stellar mass. A detailed forward modeling of the simulations to create and analyze realistic mock images and mock spectra would be needed to perform an apples-to-apples comparison with the observations at $z\sim6$. However, this is beyond the scope of the current work.

Instead, to mitigate uncertainties in the sizes of simulated galaxies and their comparison to observed galaxies, in what follows we measure the mass budget within a fixed physical aperture. Note that for simulations this is still a spherical radius, and for the observations a projected one. We set the aperture radius equal to the average half-light radius found in the $z\sim6$ JWST observations: $r_{\rm aper}=1\kpc$. We hence measure the stellar, gas and dark matter aperture masses for the $z=6$ descendant sample across $0<z<6$. Separately, for each snapshot we also measure the aperture masses for all central galaxies in the snapshot, in order to map the relative mass components as a function of redshift in narrow bins of stellar mass \citep[analogous to the work by][]{Lovell2018}.

We show the redshift evolution in $\fbar(<1\kpc)$ in the left panel of Figure~\ref{fig:fdm_aper} for the $z=6$ descendant population (black line, shaded regions). Dashed lines show the redshift evolution at fixed stellar mass, for a (color-coded) range of stellar masses. Unsurprisingly, TNG50 predicts that the baryon fraction depends strongly on stellar mass. But remarkably, Figure~\ref{fig:fdm_aper} shows that $\fbar(<1\kpc~|~M_*)$ is nearly constant from $z\sim 6$ to the present across a wide range of (fixed) stellar masses. Dash-dotted lines show the analogous estimates for the lower-resolution TNG100 simulation, implying that resolution has only a weak effect on this result. In fact, we have checked with the three different resolution levels of the same TNG50 volume that the simulation outcomes are converg{ing} in most regimes. 

For comparison, we also extract the mass components of central galaxies in the EAGLE cosmological hydrodynamical simulation \citep{Schaye2015,Crain2015}, specifically the reference model run for a volume of $100^3$ comoving Mpc$^3$ (Ref-L100N1504), using halo and subhalo catalogs that were constructed with the exact same methods as for the TNG simulations\footnote{Namely, we use a version of EAGLE that has been rewritten and analyzed exactly as the TNG data \citep[see][data release paper]{Nelson2019a}. This enables us to perform a uniform analysis on all considered simulations.}. The EAGLE simulation has similar baryonic and dark matter particle mass resolution to the TNG100 simulation and assumes a similar cosmology. In the context of this paper, the most important difference between the two simulations is the subgrid galaxy formation model, as the two use different prescriptions for, e.g., star formation, stellar feedback and black hole feedback. It is therefore interesting that -- just like the TNG simulations -- EAGLE predicts $\fbar(<1\kpc)$ depends strongly on stellar mass at fixed redshift, while $\fbar(<1\kpc~|~M_*)$ is only weakly dependent on redshift, noting that the apparent decrease from $z=1$ to $z=0$ likely reflects the fact that the mass profiles of $z\sim0$ EAGLE galaxies are affected by the resolution of the simulation for $r\lesssim 2\kpc$ \citep[see][]{Schaller2015,deGraaff2022}.

Turning to the descendants of the selected $z=6$ galaxies in TNG50, we find that for any individual descendant $\fbar(<1\kpc)$ increases rapidly toward lower redshift: while the central region is still dark matter-dominated at $z=6$ with $\fbar(<1\kpc)\sim0.25$, it is highly baryon-dominated at $z\sim2$ $(\fbar(<1\kpc)\sim0.7$). Considering the strong dependence of $\fbar$ on $M_*$, this evolution can be understood by the rapid growth of total stellar mass over time found in Figure~\ref{fig:trees}. {This $M_*$-dependence also explains the redshift-dependent ensemble scatter (i.e. the width of the contours): the sample at $z=6$ was selected in a narrow range in stellar mass, where the scatter in $\fbar$ is small (see also Figure~\ref{fig:z6_comp}). As the stellar mass distribution broadens toward lower redshift (Figure~\ref{fig:trees}), the scatter in $\fbar(<1\kpc)$ at fixed redshift also increases.  }

We also want to put the stellar-to-gas mass ratio of \citet{deGraaff2023} in the context of the TNG50 simulation. We therefore examine the stellar-to-gas mass ratio within the same 1\,kpc aperture. To plot the mass ratio $M_*/\Mgas (<1\kpc)$ in the case where $\Mgas(<1\kpc)=0$, we set a floor of $\Mgas=0.01\times M_*$. 
The right panel of Figure~\ref{fig:fdm_aper} shows that also this mass ratio depends strongly on stellar mass at fixed redshift. 
In contrast with $\fbar(<1\kpc)$, the TNG50 simulation shows a strong redshift dependence of $M_*/\Mgas$ at fixed stellar mass: the ratio is an order of magnitude higher at $z=0$ than it was at $z=6$. 

This shows that also $M_*/\Mgas(<1\kpc)\sim0.25$ for low-stellar-mass galaxies in TNG50 at $z\sim 6$ is broadly consistent with the JWST results of \citet{deGraaff2023}, who found $M_*/\Mgas(\lesssim1\kpc)\sim 0.1$. We note that, although the TNG100 simulation shows qualitatively similar trends to the high-resolution TNG50 simulation, at lower stellar masses and higher redshifts the values of $M_*/\Mgas(<1\kpc)$ can differ by up to a factor 3 between the two TNG simulations. This indicates that the precise value of $M_*/\Mgas(<1\kpc)$ depends on resolution. Therefore, it seems difficult to make a robust quantitative comparison between the observed and simulated values.

Toward lower redshift the stellar-to-gas mass ratio increases sharply for the $z=6$ descendants, and this central build-up of stellar mass coincides with the rapid growth of the total stellar mass (Figure~\ref{fig:trees}){, reminiscent of the compaction phase found in zoom-in simulations of massive galaxies at $z\sim2-4$ \citep[e.g.][]{Zolotov2015,Lapiner2023}}. At $z\lesssim2$ we find that $\Mgas(<1\kpc)\sim0$, and the mass ratio therefore plateaus at $M_*/\Mgas(\lesssim1\kpc)\sim 100$ due to our imposed gas mass floor. This likely reflects the {kinetic feedback mode from active galactic nuclei (AGN)} in the IllustrisTNG model \citep{Pillepich2018a}, in which black hole-driven winds remove gas from their surroundings.

Comparing with the EAGLE simulation, we again find qualitative agreement between the EAGLE and TNG simulations. At $z=6$ and low stellar mass the value of $M_*/\Mgas(<1\kpc)$ agrees well with that of TNG50 and therefore also agrees with the JWST observations within a factor $\sim 2$. At higher masses and lower redshift the stellar-to-gas mass ratio differs increasingly strongly from the TNG simulations, reflecting the differences in the EAGLE and IllustrisTNG galaxy formation models.

{Putting the results of Figure~\ref{fig:fdm_aper} together, we can gain insight into the physical processes behind the redshift-independence of $\fbar$ at fixed stellar mass. The strong dependence of $\fbar(<1\kpc)$ on stellar mass implies there is a $M_*$-dependent balance between the cooling of baryons on the one hand, and feedback from stellar evolution and/or AGN on the other hand. Indeed, \citet{Nelson2019} showed that this feedback in TNG50 depends strongly on stellar mass: at $z>4$, where feedback is driven primarily by stellar winds and supernovae, the mass loading factor is a factor $\approx 30$ higher for low-mass ($\sim10^8\,\Msun$) galaxies than for the highest masses ($\sim10^{11}\,\Msun$), and explains why at these high redshifts the most massive galaxies are able to rapidly form a stellar mass-dominated center. 

Toward lower redshifts the mass loading factor increases, especially for more massive galaxies, and the $M_*$-dependence for the feedback is thus reduced. Yet, despite this redshift evolution in the feedback efficiency, which implies that gas can be expelled more easily from the centers of galaxies at all masses at lower redshift, $\fbar(<1\kpc~|~M_*)$ remains constant with redshift. This can be partially explained by the fact that the cooling of baryons is also redshift-dependent. However, perhaps more importantly, the ratio of $\Mgas/M_*$ changes with redshift as well: at $z<3$ a substantial fraction of the baryonic mass is in the form of stars (even at low stellar mass), which are not affected explicitly by feedback. We note that AGN feedback, which becomes prominent at $z\lesssim2$ and high stellar masses in TNG50, therefore also plays little role in the evolution of $\fbar$, as by this time the stellar mass-dominated center has already been formed. 

It is thus likely the interplay between cooling, feedback, and central stellar mass growth that conspires to the redshift-independence of $\fbar(<1\kpc~|~M_*)$ in the TNG simulations. The fact that the EAGLE simulation with a very different subgrid model shows a qualitatively similar trend is intriguing. At $z>5$ there appears to be a slight upturn in $\fbar(<1\kpc~|~M_*)$ in the EAGLE simulation, which may reflect physical differences in the stellar feedback or cooling prescriptions with respect to the TNG model at high redshifts. }

\begin{figure*}
    \centering
    \includegraphics[width=0.75\linewidth]{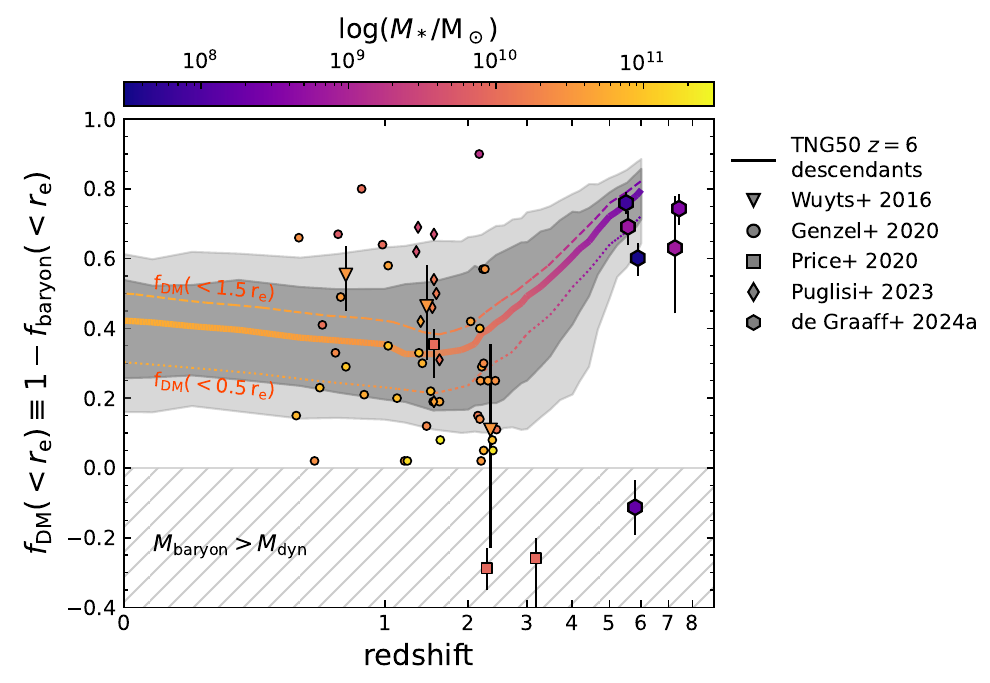}
    \caption{Redshift evolution of the dark matter fraction within the stellar half-mass radius (TNG50) and the half-light radius (observations). The dark (light) shaded regions indicate the 16-84 (5-95) percentiles of the evolution of TNG50 galaxies selected at $z=6$ with $10^8\,\Msun<M_*<10^9\,\Msun$. The thick solid line shows the median evolution of $\fdm(<\re)$ of the TNG50 galaxies; to bracket possible systematic differences in the comparison to observations, we also show the median evolution of $\fdm(<0.5\re)$ (dotted) and $\fdm(<1.5\re)$ (dashed). The median evolutionary tracks and data points from high-redshift observations are color-coded by the total stellar mass. Qualitatively, this suggests a link between observations at different mass and redshift epochs, and agreement between the TNG50 simulation and observations.  }
    \label{fig:fdm_re}
\end{figure*}

\subsection{Comparison with observations at cosmic noon}

Despite the caveat that there are substantial differences in the observed and simulated baryonic mass fractions, it is instructive to draw a qualitative picture of the redshift evolution of the mass budget within the effective radius across cosmic time. 

In Figure~\ref{fig:fdm_re} we show the dark matter fraction, $\fdm(<\re) \equiv 1-\fbar(<\re)$, as a function of redshift. For the TNG50 $z=6$ descendant sample the median evolution in $\fdm$ (solid line) is color-coded by the median stellar mass at the given redshift, and contours indicate the variation within the population. {Similar to Figure~\ref{fig:fdm_aper}, the ensemble median and scatter in $\fdm(\re)$ at fixed redshift depend strongly on the stellar mass distribution, which broadens in width and shifts to higher stellar masses toward lower redshift. The evolution in $\re$ also plays an important role, leading to a slight upturn in $\fdm(<\re)$ at $z<1$, and further increasing the ensemble scatter (see also Figure~\ref{fig:z6_comp}).} We also show results at $z\sim1-3$ from various ground-based studies that obtained dynamical masses from ionized gas kinematics \citep{Wuyts2016,Genzel2020,Price2020,Puglisi2023}. All observed data points are color-coded by the total stellar mass, typically inferred from SED modeling to photometric data. 

The effective radius for the simulated galaxies is the 3D stellar half-mass radius, which differs from the observed half-light radii, and therefore affects the comparison in $\fdm(<\re)$. 
Specifically, based on the analysis of mock JWST images of TNG50 galaxies by \citet{Costantin2023}, we estimate that the half-light radii of low-mass TNG50 galaxies at $z\sim6$ are $\approx0.3\,$dex smaller than the half-mass radii. On the other hand, at $z=3$ and $M_*\sim10^{10}\,\Msun$ the half-light radii are $\approx0.2\,$dex larger than the half mass radii. \citet{Pillepich2019} and \citet{Rodriguez2019} presented half-light radii at lower redshifts, from which we also estimate a difference of $\approx0.2\,$dex between the half-light and half-mass radii of galaxies with stellar masses $\sim10^{10-11}\,\Msun$. Therefore, we also show the median evolution for  $\fdm(<0.5\re)$ and $\fdm(<1.5\re)$ in Figure~\ref{fig:fdm_re}: the effect of the different apertures is to shift the median evolution in $\fdm$ down (up) by a factor $\sim1.4$ ($\sim1.2$).

Taking into account the large scatter in both the observations and simulations, the two appear to agree well both at intermediate and high redshifts. In detail, however, there may be systematic discrepancies, due to both the precise galaxy sample selection and the method used to infer the dark matter fraction observationally. For example,  \citet{Uebler2021} constructed and analyzed mock integral field spectroscopic observations for a sample of 7 massive ($M_*>10^{10.6}$) and highly star-forming galaxies at $z=2$ from the TNG50 simulations. By exploring mock observations at multiple viewing angles per object, they showed that there is a slight systematic bias in the inferred dark matter fraction  ($\fdm(<\re)\approx 0.34$ in TNG50 vs. $\fdm(<\re)\approx 0.20$ for real galaxies) and large scatter per object depending on the viewing angle ($f_{\rm DM, max}(<\re)-f_{\rm DM, min}(<\re) \sim 0.05-0.5$).

Furthermore, the dependence of $\fdm$ on stellar mass suggests that the observations at $z\sim2$ and $z\sim6$ may be entirely consistent with each other, if, as suggested by the TNG50 merger histories, the $z\sim6$ galaxies are progenitors of the more massive baryon-dominated systems observed at $z\sim2$. If the observed systems at $z\sim6$ can be confirmed to be dark matter-dominated (discussed in Section~\ref{sec:discussion}), these objects would form the first direct detection of the progenitors of $z\sim2$ massive galaxies.

\section{Discussion \& Conclusions}\label{sec:discussion}

We have used the cosmological simulation TNG50 to extract theoretical expectations for the stellar, gas and dark matter mass components in the central parts of high-redshift galaxies. We have explored these components' fractional contributions as a function of stellar mass and redshift. In particular, we have shown that TNG50 predicts that the central baryon fraction of galaxies is independent of redshift at fixed stellar mass: the centers of low stellar mass galaxies have always been dark matter-dominated. The central baryon fraction, is however, a strong function of stellar mass at all redshifts, and so greatly increases along the evolutionary paths of individual galaxies. The central stellar-to-gas mass ratio, on the other hand, grows strongly towards lower redshifts, even at a fixed stellar mass.

Turning to high redshifts, we compare these results to recent measurements of the mass budget at $z\sim6$ from JWST observations of galaxies of low stellar mass \citep[$M_*\sim10^{7-9}\,\Msun$;][]{deGraaff2023}.
Overall, these observational results appear in good agreement with TNG50 galaxies at $z=6$: the baryonic mass fractions are approximately equally low at low stellar masses $\fbar(<1\kpc;M_*=10^{8-9}\,\Msun)\sim0.2-0.4$. The relative mass fraction among the baryonic components (stars and gas) also agrees at least qualitatively. We find $M_*(<1\kpc)\ll \Mgas(<1kpc)$ for both the simulated and observed galaxies at $z\sim6$ and $M_*<10^9\,\Msun$. Quantitatively, on the other hand, we find that -- at face value -- the stellar-to-gas mass ratio is a factor two higher in the TNG50 simulated systems compared to the observations. However, we deem this comparison inconclusive as this mass ratio in TNG depends on the resolution of the simulation (at the $20-40\%$ level across a factor of 16 in mass resolution at $z\sim5-6$ and $M_*\sim10^{8-9}\,\Msun$).

Understanding the cold-gas content in these galaxies is of broader interest, as it speaks to the efficiency of star formation ($\Mgas/{\rm SFR}$) at high redshift. Indeed, \citet{deGraaff2023} had estimated the gas content by assuming present-epoch scaling relations for the star formation efficiency, and could therefore be underestimated by a factor 3. However, this would counter the general observed trend that the depletion time decreases toward higher redshifts \citep{Tacconi2020}. It would also imply a discrepancy in the central gas masses at the level of 1 dex with respect to TNG50.

This leaves only two plausible scenarios for the observations of the mass budget at $z\sim6$. (i) The inferred low baryon fractions imply that these high-redshift galaxies -- likely progenitors of massive galaxies at $z<2$ -- are dark matter-dominated even within $1\kpc$. (ii) Alternatively, the ionized gas kinematics, on which the observational mass budget estimates rely, could be dominated by non-gravitational motions due to outflows from feedback, most probably of stellar origin. If the resulting velocity gradients and dispersions are biased high, the inferred dynamical mass will be substantially overestimated, and hence the baryon fraction will be underestimated.

The fact that the TNG50 model is in agreement with the first scenario of a dark matter-dominated phase early in the galaxy mass assembly history shows that this first interpretation is at least \emph{plausible}. Furthermore, the simulation offers insight into the evolution of the descendants of $z=6$ galaxies with $M_*=10^{8-9}\,\Msun$, and shows that these dark matter-dominated systems are indeed progenitors of more massive ($M_*\sim10^{10}\,\Msun$) galaxies at $z\sim2$. In addition to the rapid stellar mass growth, and consistent with observations at cosmic noon, these systems are baryon-dominated in the center and have higher central stellar masses than gas masses, indicating a rapid period of stellar mass assembly in the center through star formation or accretion. Moreover, we find that these galaxies typically grow to MW-mass systems at $z\sim0$ ($M_*\sim10^{10.5}\,\Msun$). In qualitative agreement with observations \citep[e.g.][]{Martinsson2013}, the central kpc is dominated by the stellar component with minimal dark matter ($\fdm(<1\kpc)\sim0.2$), whereas the dark matter fraction within the half-mass radius is substantially higher ($\fdm(<\re)\sim0.4-0.5$). 

However, we cannot rule out at present the second scenario in which the dynamics are biased by non-gravitation motions. Further observations using integral field spectroscopy to robustly resolve the velocity fields of these high-redshift systems will be critical to distinguish between the two scenarios. 

The comparison in the mass budget between the high-redshift observations and simulations demonstrates the great potential constraining power of such observations. The TNG50 simulation makes clear predictions for the central baryon fraction and its dependence on stellar mass, and (lack thereof) on redshift. This is similarly true for the stellar-to-gas mass ratio, which also appears more sensitive to the details of the simulation (resolution, subgrid model). Therefore, systematically mapping these mass ratios as a function of stellar mass and redshift through observations would provide direct insight into galaxy mass assembly. {Current surveys with JWST that simultaneously probe the stellar population, ISM and kinematic properties of galaxies at $z>1$ across a wide range in stellar mass provide an ideal dataset for such a study \citep{Maseda2024,Eisenstein2023}.} Combined with a careful modeling analysis of observational biases by the use of mock observations from simulations, these observations will be able to place strong constraints on theoretical models and constrain the fundamental question: how efficient was star formation in the early universe?

\begin{acknowledgments}
AdG thanks Eric Rohr and Dylan Nelson for their help in an early analysis that resulted in this paper.
The IllustrisTNG simulations were undertaken with compute time awarded by the Gauss Centre for Supercomputing (GCS) under GCS Large-Scale Projects GCS-ILLU and GCS-DWAR on the GCS share of the supercomputer Hazel Hen at the High Performance Computing Center Stuttgart (HLRS), as well as on the machines of the Max Planck Computing and Data Facility (MPCDF) in Garching, Germany.
\end{acknowledgments}

% \software{astropy \citep{2013A&A...558A..33A,2018AJ....156..123A},  
%           Cloudy \citep{2013RMxAA..49..137F}, 
%           Source Extractor \citep{1996A&AS..117..393B}
%           }

\bibliography{refs}
\bibliographystyle{aasjournal}

\end{document}